\documentstyle[revtex]{aps}

\newcommand{\be}{\begin{equation}}
\newcommand{\ee}{\end{equation}}

\begin{document}
\draft

\begin{title}

Absence of Definite Scaling Laws in Raman Scattering from Fractals
\end{title}
\author{P.Benassi, O.Pilla and G.Viliani}
\begin{instit}
Dipartimento di Fisica, Universita' di Trento, i-38050 Povo, Trento, Italy
\end{instit}
\author{G.Ruocco and G.Signorelli}
\begin{instit}
Dipartimento di Fisica, Universita' di L'Aquila, via Vetoio, Coppito, I-
67100 L'Aquila, Italy
\end{instit}
\author{V.Mazzacurati}
\begin{instit}
Dipartimento STBB, Universita' di L'Aquila, Colle Maggio, I-67100 L'Aquila,
Italy
\end{instit}
\author{M.Montagna}
\begin{instit}
Dipartimento di Fisica, Universita' di Trieste, via Valerio 2, I-34127
Trieste, Italy
\end{instit}
\begin{abstract}
The frequency dependence of the Raman coupling coefficient $C(\omega)$ is
calculated numerically for square and cubic percolation clusters.
No general scaling law
in terms of the macroscopic parameters such as the fractal ($D$) and fracton
($\overline d$) dimensions is found. $C(\omega)$ is sensitive to the
microscopic structure of the clusters and depends on: site- or bond-
percolation, presence or absence of dangling bonds, scattering mechanism and
polarization, presence or absence of polarizability disorder. This situation
makes the derivation of macroscopic parameters from Raman experiments
unreliable.
\end{abstract}

The properties of fractal objects are continuing to attract much
interest from both the experimental and theoretical standpoint.
Besides being interesting on their own, fractals are thought to
be reasonably representative and relatively simple models of
important classes of real disordered materials especially as
regards the vibrational characteristics of the latter systems,
and this probably explains why their dynamics has been so
attractive to theoreticians. It is by now accepted that the
statistically self-similar, static mass- or bond-distribution of
disordered fractals (of which percolation clusters are the
prototypes) is paralleled by peculiar dynamical characteristics.
For example, the density of vibrational states follows a power
law of the type $\rho (\omega) \propto \omega ^{\overline d -1}$,
where $\overline d$ (=4/3 for percolators) is called spectral
dimension \cite{1}. Another quite general property of fractals is that
their vibrational eigenstates (fractons) are localized: if one
takes a frequency interval $\delta \omega$ and {\it averages} the
localization lengths \cite{2} of the fractons whose energy lies in the
interval $\omega \pm \delta\omega$, one finds that the average
localization length varies as $l(\omega) \propto \omega ^{-
\overline d/D}$, where $D$ is the fractal dimension. Although one
should be extremely cautious in using average quantities such as
$l(\omega)$ \cite{3}, it is not unreasonable to expect that knowledge
of $\rho(\omega)$ and $l(\omega)$ could help a lot in obtaining
information on, for example, thermal conductivity in fractals and
-possibly- in real disordered materials.
\par
Unfortunately neither $l(\omega)$ nor the low-frequency $\rho(\omega)$ are
directly and easily accessible to experiment in real disordered materials so
that
indirect measurements of these quantities might be of great help. If the
above power laws were valid, this could be achieved by measuring $\overline
d$ and $D$. In this letter we will discuss Raman scattering; there are at
least two good reasons why
people \cite{3,4,5,6,7,8,9,10,11} have concentrated on this technique: (i)
it is experimentally easy and does not require big facilities, and
(ii) since we are dealing with disordered systems, and contrary
to the case of crystals, all the eigenmodes contribute to the
scattering. The price to be paid is that the (Stokes) scattered
intensity in the harmonic approximation is given by:
\begin{equation}
I(\omega)= \{n(\omega)+1\}\rho(\omega)C(\omega)/\omega
\end{equation}
where $n(\omega)$ is the Bose-Einstein population factor for
fractons and $C(\omega)$ is an average light-vibration coupling
coefficient whose value depends on how the polarizability of the
units (atoms or molecules) which constitute the system is
modulated by the vibrational modes with frequency in the interval
$\omega \pm \delta\omega$ \cite{9}. Its $\omega$-dependence is not known a
priori. So one important question is: is it possible to cast
$C(\omega)$ in a form similar to $l(\omega)$ and $\rho(\omega)$,
i.e. $C(\omega) \propto \omega^x$, and is it possible to express
$x$ in terms of $\overline d$, $D$ and maybe other parameters
(which in the literature have been given in different times the
names of $d_\phi$ and $\sigma$)? Or, in other words, is the Raman
coupling coefficient a scaling quantity?
\par
There have been several proposals for $x$, all of which were
derived under the assumption that it is possible to determine
$C(\omega)$ by considering scaling relations of the fracton-
induced strain \cite{4,5,6,7,8,11}. In two recent papers \cite{3,9} we
argued against
such possibility on the basis of our numerical calculations of
$C(\omega)$ in two- and three-dimensional (hereafter 2D and 3D
respectively) site-percolation clusters which showed that (i)
none of the proposed $x$ expressions could fit the numerical data
in 2D and (ii) a single $x$ is not sufficient in 3D.
\par
In the present letter we extend the previous work to several
other kinds of 2D and 3D percolation structures: bond- and site-
percolation (BP and SP respectively) with and without dangling
bonds; we also investigate the effect of varying the
scattering
mechanism: full dipole-induced-dipole (FDID), DID restricted to
nearest neighbors (NNDID), bond-polarizability (BPOL) \cite{12}.
Finally we look at the effects produced by changing the bare
electric polarizabilities of the scattering units (electrical
disorder, which was not introduced in previous works).
\par
What we find is that $C(\omega) \propto \omega^x$ is a good representation
of the numerical data only for BPOL in 2D and 3D and maybe for DID in 2D; in
the
latter case however $x$ depends on the specific system considered.
$C(\omega) \propto \omega^x$ does not seem to work for DID in 3D.
Our results
also demonstrate beyond any doubt that the mechanical properties
of a system (i.e. its mass distribution and the detailed shape of its
vibrational eigenvectors) are not sufficient in general to
determine $C(\omega)$.
\par
These conclusions are at complete variance with those of a recent
letter \cite{11}, whose authors propose another scaling law for
$C(\omega)$ in terms of the mechanical parameters mentioned above
($\overline d$, $D$ and $\sigma$) and claim that this law
compares well with their numerical results relative to BP
clusters with dangling bonds \cite{13}.
\par
The numerical procedures were the same as described in Refs. 3
and 4; the clusters contained identical masses ($M=1$) and spring
constants ($K=1$), scalar elasticity was assumed. The average over
many fractons
is essential  because  the disorder-induced light scattering is an
intrinsically fluctuating quantity
and the accuracy in the determination of $C(\omega)$ depends on the number
of modes present in each frequency interval \cite{14}. The most critical
region
is
the low frequency one because of the low density of states and the big
dimensions of
the clusters needed. We overcame the problem of computation time and memory
requirements by calculating only the low frequency modes of large
clusters, while all the modes of smaller clusters were used for the
intermediate and high frequency ranges. For instance, the result for SP 2D
(two bottom traces of Fig. 1)
is obtained by averaging: all modes of
a $25 \times 25$ cluster (average over 200 realizations),
the 200 lowest energy modes of a $50 \times 50$ cluster (50 realizations),
the 400
lowest
energy modes of a $65 \times 65$ cluster (20 realizations) and the 200
 lowest energy
modes
of an $85 \times 85$ cluster (10 realizations).
\par
It is very interesting to note in Fig. 2 (3D FDID)
the important differences existing between systems with and without
dangling bonds. Dangling bonds tend to put into electromagnetic
contact regions which are dynamically uncorrelated (i.e. borders
of fjords and similar areas) and, thus, to increase the FDID
scattering mainly in the low-frequency (and low-intensity) range.
The role of electromagnetic pair interaction between units which are
spatially near but mechanically far apart has been used to explain
\cite{3,9}
the difference between FDID
and NNDID in SP. This effect is not present in BPOL where the examined
systems have very similar behaviors and follow the $\omega^x$ law in a very
broad range both in 2D ($x \approx 1.3$) and in 3D (Fig. 3, $x \approx
1.6$).
The authors of ref. 11 found no
relevant difference between BP FDID and BP NNDID and we find the same
result; but in the case of BP clusters what is meaningful is the comparison
between FDID and BPOL. In fact, while SP BPOL and SP NNDID are {\it the same
thing}, this is
not true for BP: here nearest neighbor units can be found which are not
bound, so that most of the above mentioned electromagnetic pair interaction
is already present in NNDID.
\par
We also observed tangible
though minor differences between VV and VH
polarizations in FDID for all kinds of clusters.
\par
All these considerations show that there is an
important local contribution to $C(\omega)$ because globally
equal but locally different clusters produce different
$C(\omega)$'s. This suggests that simple scaling arguments are
unlikely to apply to this quantity. This view is strengthened
by inspection of Fig. 4 which is relative to clusters whose units
have been randomly assigned the bare polarizability values
$\alpha_1$ and $\alpha_2$ with equal probabilities. The clusters
which produce the three traces of Fig. 4 are the same, yet they yield
different values of $x$, i.e. different "scaling laws" are found
for {\it the same} mechanical system with the same scattering
mechanism. Similar results are found for BPOL: introduction of the
same electrical disorders as in Fig. 4 changes the slope from $x\approx 1.3$
to $x \approx 1.9$, a value very close to that of homogeneous systems
\cite{15}.
\par
The reason why $C(\omega)$ depends so markedly on the microscopic
electrical and mechanical details of the system is to be found in
the very nature of Raman scattering. Disordered systems
scatter macroscopically because disorder does not allow the waves
scattered by different units to interfere so as to cancel exactly the
scattered amplitude as happens in crystals. In other
words, Raman scattering is due to fluctuations, i.e. it is the
mean square deviation of a zero-average quantity \cite{14}. It is
not surprising that what is not exactly annulled by the
interference process (i.e. the intensity at a given frequency)
depends on the mechanical and electrical detailed structure and
on how the respective disorders correlate.
\par
Raman scattering has been used in recent literature in order to
determine the mechanical parameters $\overline d$, $D$ and
$d_\phi$ or $\sigma$ of real systems by comparing the assumed
functional form of the exponent $x(\overline d,D,d_\phi,\sigma)$
to the "observed" $C(\omega)$ \cite{16}.
On the basis of the present results it appears that such
procedure is devoid of any significance.
\par
Whether or not it will be possible to derive reliable expressions
for $C(\omega)$ is of course still an open question; the
numerical results seem to suggest that a power law works in some
cases (BPOL) and two different slopes are present in other cases, while
in a few other instances this functional form seems to be pretty
inadequate. Even in BPOL we find $x_{2D}<x_{3D}$, contrary to what is
 expected
on the basis of the scaling models proposed so far \cite{4,8}.

\figure{ FDID, VV-polarized Raman coupling coefficient $C(\omega)$
for
2D clusters at percolation threshold. Full triangles: BP; open triangles:
BP without
dangling bonds; full circles: SP; squares: SP without dangling bonds. The
bars indicate the rms deviation. }

\figure{Same as Fig. 1, but 3D.}

\figure{ Same as Fig. 1, but 3D BPOL. }

\figure{ Effect of polarizability disorder on FDID, VV-polarized
$C(\omega)$ for 2D BP clusters; each unit is randomly assigned a bare
polarizability $\alpha_1$ or $\alpha_2$ with 50\% probability. Triangles: no
electrical disorder ($\alpha_1$=$\alpha_2$=1); squares: $\alpha_1=0.5$,
$\alpha_2 =1.5$; circles: $\alpha_1=0$, $\alpha_2=2$.}


\begin{references}
\bibitem{1}S. Alexander and R. Orbach, J. Phys. (Paris) Lett. 43, L-
625 (1982).

\bibitem{2} As discussed in Refs. 3 and 9-11, there are different ways of
defining the localization length which require different
averaging intervals $\delta\omega$ to produce a reasonably smooth
function $l(\omega)$.

\bibitem{3} M. Montagna, O. Pilla, G. Viliani, V. Mazzacurati, G.
Ruocco and G. Signorelli, Phys. Rev. Lett. 65, 1136 (1990).

\bibitem{4} A. Boukenter, B. Champagnon, E. Duval, J. Dumas, J.F.
Quinon and J. Serughetti, Phys. Rev. Lett. 57, 2391 (1986).

\bibitem{5} T. Keyes and T. Ohtsuki, Phys. Rev. Lett. 59, 603 (1987).

\bibitem{6} E. Duval, G. Mariotto, M. Montagna, O. Pilla, G. Viliani
and M. Barland, Europhys. Lett. 3, 33 (1987).

\bibitem{7} A. Fontana, F. Rocca and M.P. Fontana, Phys. Rev. Lett. 58,
503 (1987).

\bibitem{8} Y. Tsujimi, E. Courtens, J. Pelous and R. Vacher, Phys.
Rev. Lett. 60, 2757 (1988).

\bibitem{9} V. Mazzacurati, M. Montagna, O. Pilla, G. Viliani, G.
Ruocco and G. Signorelli, Phys. Rev. B 45, 2126 (1992).

\bibitem{10} O. Pilla, G.Viliani, M. Montagna, V. Mazzacurati, G. Ruocco
and G. Signorelli, Philos. Mag. B 65, 243 (1992).

\bibitem{11} E. Stoll, M. Kolb and E. Courtens, Phys. Rev. Lett. 68,
2472 (1992).

\bibitem{12} See ref. 9 for a deeper discussion of the DID and BPOL
mechanisms.

\bibitem{13} The authors of Ref. 11 assert that BP clusters should always
be preferred to SP in checking scaling properties; this
proclamation is curious enough because from their Fig. 4 it
appears that 3D SP follows their eq. (3) much more closely and in
a broader frequency range than 2D and 3D BP, which look rather
roundish (see also Figs. 1 and 2 of the present letter). Their
conviction might be explained by the fact that fitting 2D SP,
which has $x \approx 1$ (see Refs. 3 and 9), would require an
unpalatable $\sigma \approx 1.5$. See Refs. 3 and 9 also for a
different explanation of the presence of two slopes in 3D SP. In
any case, for the reasons explained in the text, we believe that
there is not much sense in eq. (3) of Ref. 11; among others, in
the limit $\overline d = D = 3$, $\sigma=1$ (i.e. a homogeneous
connected system) it leaves us with the eccentric value $x=-1$ instead of
$x=2$.

\bibitem{14} M. Montagna, P. Benassi, W. Frizzera, O. Pilla, G. Viliani,
V. Mazzacurati, G. Ruocco and G. Signorelli, Physica B (1992), in press.

\bibitem{15} P. Benassi, O. Pilla, V. Mazzacurati, M. Montagna, G.
Ruocco and G. Signorelli, Phys. Rev. B 44, 11734 (1991),
discussed the effect of various kinds of electrical disorder on
$C(\omega)$ in topologically perfect crystals, and found that it
strongly depends on the site polarizability distribution.

\bibitem{16} Of course, what is actually observed is the product
$C(\omega)\rho(\omega)$, eq. (1), and in order to extract
$C(\omega)$ one must either make assumptions on $\rho(\omega)$ as
described at the beginning of this paper, or measure it by
neutron scattering, but the range of frequencies where real solids are
expected to be fractal is very low.
\end{references}
\end{document}